\documentclass[aps,prx,twocolumn]{revtex4-1} 
\usepackage{graphicx}
\usepackage{amsmath,amsfonts,amssymb}
\usepackage{bm}
\usepackage{color}

\usepackage{xcolor}

\usepackage{ulem}

\usepackage{soul}

\begin{document}

\newcommand{\matteo}[1]{{\color{green} #1}}
\newcommand{\commento}[1]{{\color{red} #1}}


\title{Active Matter at high density: velocity distribution and kinetic temperature}

\author{Lorenzo Caprini}
\email{lorenzo.caprini@gssi.it}
\affiliation{Universit\'a di Camerino, Dipartimento di Fisica, Via Madonna delle Carceri, I-62032 Camerino, Italy}

\author{Umberto Marini Bettolo Marconi}
\affiliation{Universit\'a di Camerino, Dipartimento di Fisica, Via Madonna delle Carceri, I-62032 Camerino, Italy}



\begin{abstract}
We consider the solid or hexatic non-equilibrium phases of an interacting two-dimensional system of Active Brownian Particles at high density
and investigate numerically and theoretically the properties of the velocity distribution function and the associated kinetic temperature. 
We obtain approximate analytical predictions for the shape of the velocity distribution and find a transition from a Mexican-hat-like to a Gaussian-like distribution as the persistence time of the active force changes from the small to the large persistence regime.
Through a detailed numerical and theoretical analysis of the single-particle velocity variance, we report an exact analytical expression for the kinetic temperature of dense spherical self-propelled particles that holds also in the non-equilibrium regimes with large persistence times and discuss its range of validity.
\end{abstract}

\maketitle

\newcommand{\betadef}{\frac{1}{\tau}}
\newcommand{\alphadef}{\frac{\omega_q^2}{\gamma}}
\newcommand{\br}{{\bf r}}
\newcommand{\bu}{{\bf u}}
\newcommand{\bR}{{\bf x}}
\newcommand{\bRz}{{\bf x}^0}
\newcommand{\bk}{{ \bf k}}
\newcommand{\bx}{{ \bf x}}
\newcommand{\vv}{{\bf v}}
\newcommand{\nb}{{\bf n}}
\newcommand{\mb}{{\bf m}}
\newcommand{\bq}{{\bf q}}
\newcommand{\rb}{{\bar r}}

\newcommand{\eeta}{\boldsymbol{\eta}}
\newcommand{\xxi}{\boldsymbol{\xi}}

\section{Introduction}


Active Matter at high density is becoming a subject of great interest since it plays a crucial role to understand a broad range of biological systems~\cite{bechinger2016active, marchetti2013hydrodynamics, fodor2018statistical, gompper20202020}, such as cell monolayers and living tissues.
Experimental observations realized with cell monolayers reveal large-scale collective motion, like swirls and velocity alignment~\cite{henkes2020dense, sepulveda2013collective, garcia2015physics, basan2013alignment, nava2020modelling}.
In the spirit of minimal modeling, these systems have been recently modeled using high-density interacting Active Brownian Particles (ABP)~\cite{sarkar2020minimal, henkes2020dense}, thus, modeling the complex cells and cell-substrate interactions through steric interaction and self-propulsion.
Depending on their density, interacting systems of ABP display a variegate phenomenology.
In particular, at moderate packing fractions, a non-equilibrium phase-coexistence known as Motility-Induce Phase-Separation (MIPS)~\cite{fily2012athermal, buttinoni2013dynamical, cates2015motility,  gonnella2015motility, ma2020dynamic} occurs even in the absence of attractive interactions~\cite{caprini2020spontaneous, stenhammar2014phase, siebert2017phase, ginot2018aggregation, solon2018generalized, van2019interrupted, mallory2019activity, grauer2020swarm, jose2020phase, chiarantoni2020work}.
Depending on the active force and the packing fraction, ABP can also attain homogeneous configurations, such as active liquid, hexatic and solid phases \cite{bialke2012crystallization, menzel2013traveling, briand2018spontaneously, klamser2018thermodynamic, digregorio2018full, de2020phase, paliwal2020role}.
With respect to equilibrium systems of Brownian colloids, the active liquid-hexatic and hexatic-solid transitions are shifted towards larger values of the packing fractions and the hexatic phase occurs for a broad range of parameters~\cite{digregorio2018full, cugliandolo2017phase, digregorio2019clustering, caprini2020hidden}.
Moreover, the dense phases of Interacting ABP display a plethora of dynamical phenomena making them quite different as compared to passive dense phases. In particular, the particle velocities spontaneously form ordered domains even in the absence of explicit alignment interactions in phase-separated configurations~\cite{caprini2020spontaneous} and in active liquid, hexatic and solid phases~\cite{caprini2020hidden} and give rise to fascinating intermittency phenomena~\cite{caprini2020hidden, mandal2020extreme}. 
The spontaneous alignment mechanism makes the ABP models suitable to describe the behavior of cell monolayers.

For systems of interacting ABP, several authors,
searching for an extension of equilibrium thermodynamic concepts  introduced an effective ``temperature''
in the study of non-equilibrium systems of self-propelled particles.
Following ideas from glassy systems, several authors introduce the dynamical effective temperature by the ratio between the mean-square-displacement and the time-integrated linear response function due to small perturbation, in the context of active disks~\cite{palacci2010sedimentation, berthier2013non, levis2015single, preisler2016configurational, nandi2017nonequilibrium, szamel2017evaluating, nandi2018effective, cugliandolo2019effective, dal2019linear, petrelli2020effective}, dumbells~\cite{suma2014dynamics, petrelli2018active}, polymers~\cite{loi2011non}, looking both at active homogeneous (liquid, hexatic and solid) configurations and phase-separated regimes.
In the homogeneous case, the effective temperature increases as the propulsive speed increases and decreases as the packing fraction grows. Instead in the inhomogeneous case, i.e. when MIPS occurs, the net distinction between the populations of the two coexistent phases, e.g. slow particles in the dense clusters and fast particles in the disordered phase, allows us to introduce two distinct effective temperatures~\cite{petrelli2020effective}.
 Mandal et. al.~\cite{mandal2019motility} focused on the kinetic temperature, i.e. the variance of the velocity distribution, for underdamped self-propelled particles, outlining that the temperatures in the two coexisting phases of MIPS are different.
An alternative definition of the active temperature has been also proposed in the context of stochastic thermodynamics to generalize the Clausius relation to active systems, 
a program requiring the introduction of a space-dependent temperature that depends on the potential itself~\cite{marconi2017heat, caprini2019activityinduced, szamel2014self}.


In this work, we shall not discuss the concept of temperature in non-equilibrium active systems~\cite{puglisi2017temperature}  
intended as an observable satisfying well defined thermodynamic relations, an issue still matter of debate, but focus on the kinetic temperature of self-propelled particles. We find its exact analytical expression as a function of the model parameters for dense homogeneous configurations in the non-equilibrium active solid and hexatic phases. We also
obtain the ABP single-particle velocity distribution in these highly packed configurations 
as the persistence time of the self-propulsion varies.
The shape of this distribution obtained via numerical simulations is compared
with theoretical predictions both in the small and the large persistence regimes. We find a crossover between the two regimes 
which manifests itself in a qualitative change in the shape of the velocity distribution.

The manuscript is organized as follows: in Sec.~\ref{Sec:Model}, we introduce the model, while, in Sec.~\ref{Sec:velocitydynamics}, we report the velocity dynamics representing the starting point of our theoretical approach. Numerical and theoretical results of single-particle velocity distributions are shown in Sec.~\ref{Sec:velProb}, while the analysis of the first moments of the distribution and the discussion about the kinetic temperature are reported in Sec.~\ref{Sec:kineticTemperature}.
Finally, we report some discussions and conclusions in the final section.

\section{Interacting self-propelled particles}\label{Sec:InteractingABPparticles}\label{Sec:Model}

We consider a two-dimensional system of $N$ self-propelled disks, described by the Active Brownian Particles (ABP) model, where inertial and hydrodynamic effects are neglected.
The position, $\mathbf{x}_i$, of each disk evolves by the following stochastic differential equation:
\begin{equation}
\label{eq:x_dynamics_ABP}
\gamma\dot{\mathbf{x}}_i = \mathbf{F}_i + \mathbf{f}^a_i \,.
\end{equation}
The constant $\gamma$ is the solvent friction while we do not take into account the thermal diffusivity
since for several experimental active particle systems~\cite{bechinger2016active}
 is some orders of magnitude smaller than the diffusivity associated with the self-propulsion force, $\mathbf{f}^a_i$. According to the popular ABP model, $\mathbf{f}^a_i$ is a time-dependent force given by the equation:
\begin{equation}
\mathbf{f}^a_i=\gamma v_0 \mathbf{n}_i \,,
\end{equation}
where $v_0$ is the constant modulus of the swim velocity induced by $\mathbf{f}^a_i$ and $\mathbf{n}_i$ is the orientation vector of components $(\cos\theta_i, \sin\theta_i)$ evolving through a stochastic process.
In particular, the orientational angle, $\theta_i$, performs angular diffusion:
\begin{equation}
\label{eq:theta_dynamics}
\dot{\theta}_i= \sqrt{2D_r} \,\xi_i  \,,
\end{equation}
where $\xi_i$ is a white noise with unit variance and zero average and $D_r$ is the rotational diffusion coefficient.
We remark that the inverse of $D_r$ defines the correlation-time of the active force, namely $\tau=1/D_r$~\cite{farage2015effective}, which will be assumed as a control parameter in the numerical study performed in this manuscript.

The term $\mathbf{F}_i$ represents the repulsive force between particles due to steric interactions.
In particular, $\mathbf{F}_i=-\nabla_i U_{tot}$ where the potential, $U_{tot}$, can be expressed as $U_{tot}= \sum_{i<j} U(|{\mathbf x}_{ij}|)$, with ${\mathbf x}_{ij}=\mathbf{x}_i -\mathbf{x}_j$.
We choose $U(r)$ as a shifted, truncated Lennard Jones Potential~\cite{redner2013structure, caprini2020spontaneous}:
\begin{equation}
\label{eq:potentialshape}
U(r)=
4\epsilon\left[\left(\dfrac{\sigma}{r}\right)^{12}- \left(\dfrac{\sigma}{r}\right)^{6}\right] + \epsilon,  \quad \,\, r\leq2^{1/6}\sigma
\end{equation}
and zero for $r>2^{1/6}\sigma$.
The constant $\epsilon$ is the typical energy scale of the interactions, while $\sigma$ is the nominal particle diameter.
The short-range nature of the potential allows us to consider only the force contributions of first-neighboring particles even in the very packed configurations considered in this paper.
Both $\epsilon$ and $\sigma$ are set to one for numerical convenience.


We focus on high density regimes exploring the homogeneous aggregation phases of self-propelled particles.
In particular, we fix $v_0=50$ and the packing fraction, $\phi=N/L^2 \sigma^2/4$, to the value $1.1$, where
 the system attains active solid or hexatic configurations without showing density inhomogeneities~\cite{caprini2020hidden}.
In particular, the hexatic-solid transition is controlled by $\tau$ and occurs approximatively at $\tau=0.1$.
Under these conditions, we study the single-particle velocity distribution varying $\tau$ and its moments.
We can distinguish between two regimes~\cite{caprini2020hidden}: i) the small persistence regime where $\tau< U''(\bar{r})/\gamma$ and ii) the large persistence regime where $\tau > U''(\bar{r})/\gamma$, being $\bar{r}$ the average distance between neighboring particles that is fixed by the density in any homogeneous configurations.
In the case i), the self-propulsion $\mathbf{f}^a_i$ is the fastest degree of freedom:
in this regime, the persistence time, $\tau$, is smaller than the typical time of the potential $U''(\bar{r})/\gamma$, so that the behavior of ABP resembles that of passive Brownian particles and the $\mathbf{x}_i$ just 
display oscillations around their equilibrium positions.
Considering the structural properties of the system, this regime is indistinguishable from the passive solid-state.
In case ii), the evolution of $\mathbf{f}^a_i$ plays a relevant role and affects the dynamics of $\mathbf{x}_i$, manifesting 
itself in several dynamical anomalies~\cite{caprini2020spontaneous, caprini2020hidden} due to the intrinsic non-equilibrium nature of active models.

\section{The velocity dynamics}\label{Sec:velocitydynamics}

As already reported in~\cite{caprini2020spontaneous, caprini2020hidden, caprini2020time},
the study of the velocity dynamics reveals the existence of hidden collective behavior of self-propelled particles at high density in the regime of large persistence times.
Nevertheless, many single-particle properties, such as the velocity distribution and its moments, have not been yet explored.

Following~\cite{caprini2020spontaneous}, we eliminate
$\mathbf{f}^a_i$ in favor of $\mathbf{v}_i=\dot{\mathbf{x}}_i$, i.e. the velocity of the particle, which does not coincide with the swim velocity, $v_0 \mathbf{n}_i$, since the modulus of ${\mathbf{v}}_i$ is not fixed and its orientation 
is not parallel to $\mathbf{n}_i$. This statement is true when particles interact and, thus, at high densities, in particular.
Transforming the dynamics from the variables ($\mathbf{x}_i, \mathbf{f}^a_i$) to the new variables ($\mathbf{x}_i, \mathbf{v}_i$) (without any approximations), the equations of motion read:
\begin{subequations}
\label{eq:dynamics_xv}
\begin{align}
\label{eq:xv_dynamics_MIPS_varx}
\dot{\mathbf{x}}_i &= \mathbf{v}_i \\
\label{eq:xv_dynamics_MIPS}
\tau\gamma\dot{\mathbf{v}}_i &= - \gamma\sum_{j=1}^N{\boldsymbol{\Gamma}}_{ij}({\mathbf x}_{i}-{\mathbf x}_{j}) \mathbf{v}_j + \mathbf{F}_i +  \tau\gamma\mathbf{k}_i 
\end{align}
\end{subequations}
where  each $\boldsymbol{\Gamma}_{ij}$ is two-dimensional matrix with components
\begin{equation}
\label{eq:Gammadefinition_MIPS}
\Gamma_{ij}^{\alpha \beta}({\mathbf r}_{ij}) = \delta_{ij}\delta_{\alpha \beta} + \frac{\tau}{ \gamma} \nabla_{i\alpha} \nabla_{j \beta} U(|{\mathbf r}_{ij}|) \,.
\end{equation}
Greek indices are used to denote the spatial components $\alpha, \beta=x, y$ while Latin indices identify the particle number $i,j=1, ..., N$.
Finally, the term $\mathbf{k}_i$ is a noise vector that reads:
\begin{equation}
\mathbf{k}_i 
=v_0 \sqrt{\frac{2}{\tau}} \,\boldsymbol{\xi}_i \times \frac{\gamma\mathbf{v}_i+ \nabla_i U_{tot}}{\gamma v_0} \,,
\end{equation}
where $\boldsymbol{\xi}_i$ is a vector with components $(0,0, \xi_i)$ and normal to the plane of motion, $(x, y, 0)$.
The vector $\mathbf{k}_i$ is a multiplicative noise depending both on $\mathbf{v}_i$ and $\mathbf{F}_i$
and is perpendicular to $\mathbf{n}_i$, i.e. the orientation of the active force. Its amplitude scales simply as $\sim v_0 \sqrt{2/\tau}$ since $\mathbf{n}_i$ is a unit vector.

The dynamics~\eqref{eq:xv_dynamics_MIPS} resembles the evolution of underdamped passive particles which are out from equilibrium because of the occurrence of space-dependent friction forces (e.g. the diagonal terms of the matrix $\boldsymbol{\Gamma}$) and effective forces depending on positions and velocities of neighboring particles (e.g. the non-diagonal terms of $\boldsymbol{\Gamma}$).
Eq.~\eqref{eq:xv_dynamics_MIPS} resembles the dynamics of the Active Ornstein-Uhlenbeck particle (AOUP)  ~\cite{marconi2016velocity, fodor2016far, wittmann2019pressure, berthier2019glassy, caprini2019active, woillez2020active, maggi2017memory, fily2019self, dabelow2019irreversibility}, 
an alternative model used to study the behavior of self-propelled particles.
Upon a suitable mapping of the self-propulsion parameters~\cite{caprini2019comparative, das2018confined}, the difference between AOUP and ABP dynamics is represented by the noise term $\mathbf{k}_i$, which in the former is a white noise vector with independent components~\cite{marconi2016velocity, fodor2016far}.

\section{Probability distribution function of the velocity}\label{Sec:velProb}

\begin{figure*}[t]
\centering
\includegraphics[width=0.99\textwidth,clip=true]{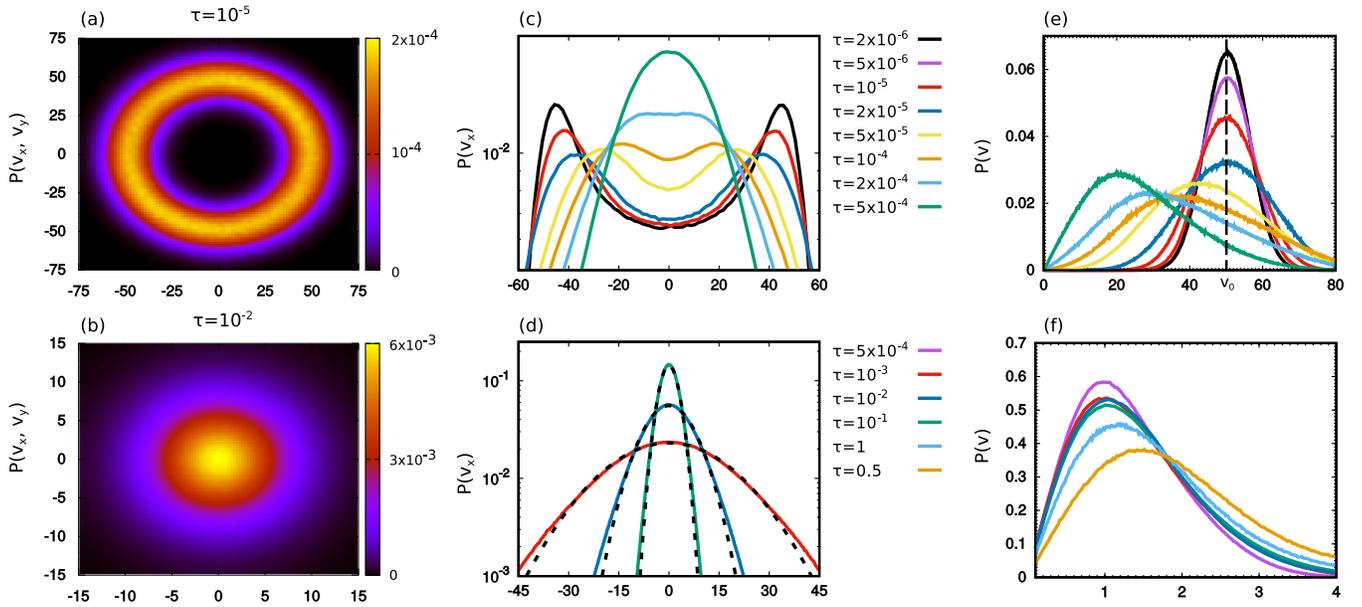}
\caption{Probability distribution of the velocity. Panels (a) and (b): map of the two-dimensional probability distribution function, $p(v_x, v_y)$ for two different values of $\tau=10^{-5}$ (panel (a)) and $\tau=10^{-2}$ (panel (b)).
 Panels (c) and (d): marginal probability distribution function, $p(v_x)$, for several values of $\tau$ (colored lines).
The dashed black lines in panel (d) are obtained by numerical fits, obtained with Gaussian distributions.
Panels (e) and (f) report the probability distribution of the velocity modulus, $p(|\mathbf{v}|=v)$, for different values of $\tau$.
In particular, in panel (f), we show $p(v)$ rescaled by $\tau^{2/5}$.
Panels (c), (e), and (d), (f) share the captions. 
Numerical simulations are obtained with $v_0=50$, $\epsilon=\sigma=1=\gamma=1$.
}
\label{fig:velocitypdf}
\end{figure*}

We numerically study the distribution of the velocity in the steady-state to evaluate the effect of the persistent time, $\tau$.
As illustrated in Fig.~\ref{fig:velocitypdf} (a) and (b), the distinction between large and small persistence regimes produces different shapes in the probability distribution function of the velocity, $p(v_x, v_y)$.
In the small-$\tau$ configurations shown in panel (a) [case i)], $p(v_x, v_y)$ has a pronounced non-Gaussian shape: the probability of finding a particle with $v\approx 0$ is negligible and the velocity of the particles is peaked around a circular crown with radius $\approx v_0$.
Instead, in the large-$\tau$ configurations reported in panel (b) [case ii)], $p(v_x, v_y)$ presents a Gaussian-like shape quite similar to the case of passive Brownian particles.
Intuitively, in case i), the self-propulsion changes rapidly without producing any appreciable change in the particle positions  
giving rise only to very small fluctuations. The net steric force exerted by the neighboring particles on a tagged particle almost cancels out and has little effect on the particle velocity, that in practice only experiences the influence of the active force.
This explains why the distribution of  $\mathbf{v}$ 
is very similar to the one of $\mathbf{f}^a$.
On the contrary, in the large $\tau$ regime, this is no longer true. 
The direction of the active force before appreciably changing need an interval $\sim\tau$ 
much larger than the relaxation time associated with the interparticle potential, $\tau_p=U''(\bar{r})/\gamma$. 
Now, the resultant of the active and steric forces nearly vanishes and (see Eq.\eqref{eq:x_dynamics_ABP}), as a consequence,
the average velocity is almost zero and the single-particle kinetic energy decreases. 

The transition from regime i) to regime ii) is quantitatively evaluated in Fig.~\ref{fig:velocitypdf}~(c)-(d) where the marginal probability distribution of the velocity along one component, $p(v_x)=\int dv_y p(v_x, v_y)$, is studied for different values of $\tau$.
For smaller values of $\tau$, $p(v_x)$ displays two symmetric peaks near $v_x \approx v_0$.
When $\tau$ grows, the peaks shift towards smaller values of $v_x$ and their heights decrease with respect to the $p(v_x)$ value around the origin.
For $\tau\gtrsim 2\times 10^{-4}$, the two peaks merge and the bimodality of the distribution is suppressed, while for further $\tau$-values a single pronounced peak placed at $v_x\approx 0$ occurs.
The deviation between $p(v_x)$ and a Gaussian distribution is not very pronounced as revealed in Fig.~\ref{fig:velocitypdf}~(d), as shown by the comparison with the best Gaussian fit.
To provide another perspective, we study the probability distribution of the velocity modulus, $p(v=|\mathbf{v}|)$, in Fig.~\ref{fig:velocitypdf}~(e)-(f).
In the small-$\tau$ regime (panel (f)), the distribution is peaked around $v=v_0$ displaying a quite symmetric shape fairly described by a Gaussian centered in $v=v_0$:
\begin{equation}
\label{eq:distribution_taularge}
p(v) \approx \mathcal{N} v\exp{\left[- \frac{\alpha}{2}\left(v -v_0 \right)^2\right]} 
\end{equation}
where $\mathcal{N}$ is a normalization factor and $\alpha$ a parameter that satisfies, $\langle v^2 \rangle =\langle v\rangle^2 + 3\alpha/2$.
We observe that $p(v)$ becomes narrow as $\tau$ grows, in the small-$\tau$ regime, but, for further values of $\tau$, the peak of the distribution shifts towards smaller values. 
After a crossover regime, occurring for intermediate values of $\tau$, the distribution $p(v)$ approaches a Gaussian-like shape such that
\begin{equation}
\label{eq:distribution_tausmall}
p(v)\approx v \exp{( - v^2/\beta)} \,,
\end{equation} 
where $1/\beta=\langle v^2\rangle$, consistently with the observations of panel (d).
Interestingly, in panel (f), we show that the $v$-distribution collapses for $v\to v \tau^{2/5}$ for a large range of $\tau$ between $(10^{-3},10^{-1})$.
Thus, $1/\beta$, which plays the role of an effective temperature, decreases as $\tau$ is enlarged.
We remark that the scaling of $p(v)$ ceases to hold for values $\tau \gtrsim 10^{-1}$ when the homogeneous active solid phase
breaks down.
These observations will be clarified in the next theoretical sections.

\subsection{Theoretical predictions}

From the set of stochastic equations~\eqref{eq:dynamics_xv}, we derive the Fokker-Planck equation for the probability distribution function, $p=p(\{ x\}, \{v\})$ (where the symbol $\{ \cdot \}$ has been introduced to denote all the space-components of the $N$  particles):
\begin{equation}
\label{eq:app_FokkerPlanck}
\begin{aligned}
\frac{\partial}{\partial t} p =& -\mathbf{v}_i \cdot \nabla_{\mathbf{x}_i} p + \frac{1}{\tau} \left( \mathcal{I}_{ij} + \frac{\tau}{\gamma} \nabla_{\mathbf{x}_i} \nabla_{\mathbf{x}_j} U  \right) \nabla_{\mathbf{v}_i} \cdot\left( \mathbf{v}_j p\right) \\
&+ \frac{\nabla_{\mathbf{x}_i} U}{\tau \gamma}\cdot\nabla_{\mathbf{v}_i} p + \frac{v_0^2}{\tau}\nabla_{\mathbf{v}_i} \nabla_{\mathbf{v}_j} \left(\mathbf{\mathcal{D}}_{ij} p\right)\,,
\end{aligned}
\end{equation}
where $\mathcal{I}_{ij}$ is the identity matrix and each element $\mathbf{\mathcal{D}}_{ij}$ is a $2\times2$ symmetric matrix of the form: 
\begin{equation*}
\begin{aligned}
\mathbf{\mathcal{D}}_{ij} &= \delta_{ij} \begin{bmatrix} 
n^2_{y_i} \quad& -n_{x_i} n_{y_i} \quad&0\\
-n_{x_i} n_{y_i} \quad& n^2_{x_i} \quad&0 \\
0\quad&0\quad&0  
\end{bmatrix} \,.
\end{aligned}
\end{equation*}
We remark that each $\mathbf{\mathcal{D}}_{ij}$ is a non-diagonal matrix as a consequence of the complex noise structure in Eq.~\eqref{eq:dynamics_xv} and of the fact that $n_{x (y)}$ is a function of the particle velocity and position through Eq.~\eqref{eq:x_dynamics_ABP}.
We point out that Eq.~\eqref{eq:app_FokkerPlanck} has the same form as the AOUP Fokker-Planck equation except for the diffusion-like term (i.e. the term containing   $\mathbf{\mathcal{D}}_{ij}$ in Eq.~\eqref{eq:app_FokkerPlanck}). 
However, in the AOUP equation, the non-diagonal matrix $\mathcal{D}_{ij}$ is replaced by a diagonal one, $\tilde{\mathcal{D}}_{ij}$, with components 
$$
\tilde{\mathcal{D}}_{ij} =\delta_{ij}\begin{bmatrix} 
1 \quad& 0 \quad&0\\
0 \quad& 1 \quad&0 \\
0\quad&0\quad&0 
\end{bmatrix} \,.
$$
We observe that $\tilde{\mathcal{D}}_{ij}$ can be obtained from $\mathcal{D}_{ij}$ just by replacing $n_{x(y)}^2$ and $n_x n_y$ by their averages, i.e. $\langle n_{x(y)}^2\rangle=1/2$ and $\langle n_x n_y\rangle=0$, respectively (with the addition of an extra factor 2 needed for consistency between the parameters of the two models~\cite{caprini2019comparative}).
We remark that, even in the simplified AOUP case, the solutions of Eq.~\eqref{eq:app_FokkerPlanck} for $\tau> 0$ and generic potential are only known in the regime of small persistence, in particular, as an expansion in powers of $\tau\gamma$ around a Gaussian distribution~\cite{fodor2016far, martin2020statistical}.
On the contrary, in the case of interacting ABP, there are neither asymptotic nor approximated results for the probability distribution function of the velocity.

\begin{figure}[t]
\centering
\includegraphics[width=0.85\columnwidth,clip=true]{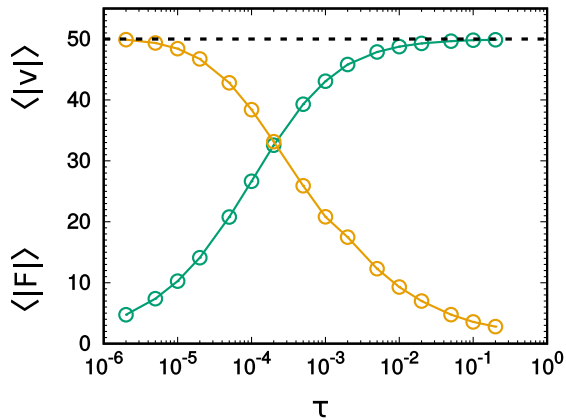}
\caption{
$\langle |\mathbf{v}| \rangle$ (yellow data) and $\langle |\mathbf{F}|\rangle$ (green data) as a function of $\tau$.
The black dashed line is drawn in correspondence of $v_0$ that is chosen as $v_0=50$ in the numerical simulations. 
The other parameters are $\epsilon=\sigma=\gamma=1$.
}
\label{fig:PotVel}
\end{figure}

Being the general solution of Eq.~\eqref{eq:app_FokkerPlanck} unknown, we will employ suitable approximations supported by numerical observations.
In Fig.~\ref{fig:PotVel}, we report $\langle|\mathbf{F}|\rangle$ and $\langle|\mathbf{v}|\rangle$ as a function of $\tau$.
In the small-persistence regime, $v_0 \approx\langle|\mathbf{v}|\rangle \gg \langle|\mathbf{F}|\rangle$ while in the large-persistence regime the opposite relation holds, namely $v_0 \approx \langle|\mathbf{F}|\rangle \gg \langle|\mathbf{v}|\rangle$, confirming the physical explanation mentioned before.
Such observation will be crucial in the following to derive approximate analytical solutions of $p(v)$.

\subsubsection{Large persistence regime}


Using the observation $v_0 \approx \langle|\mathbf{F}|\rangle \gg \langle|\mathbf{v}|\rangle$, holding in the large persistence regime, we have $\mathbf{n}_i \approx \nabla_i U$, and the diffusive matrix can be approximated as follows:
\begin{equation*}
\begin{aligned}
\mathbf{\mathcal{D}}_{ij} &\approx \delta_{ij}\frac{1}{v_0^2\gamma^2} \begin{bmatrix} 
\left(\nabla_{y_i} U\right)^2 \quad& -\nabla_{y_i} U \nabla_{x_i} U \quad&0\\
-\nabla_{y_i} U \nabla_{x_i} U \quad& \left(\nabla_{x_i} U\right)^2 \quad&0 \\
0\quad&0\quad&0 
\end{bmatrix} \,.
\end{aligned}
\end{equation*}
In the active solid phases, where the defects of the crystalline arrangement are negligible, the sum of the forces exerted by neighboring particles cancel out and we can assume $\nabla_{x_i(y_i)}U =0$. 
In other words, our approximation consists in replacing $\nabla_{x_i(y_i)}U$ with $\langle \nabla_{x_i(y_i)}U\rangle=0$. 
This is true only in the high-density regime where we can approximate the gradient of the potential expanding the distance between neighboring particles around $\bar{r}$.
Since $\langle \nabla_{x_i}U \nabla_{y_i}U\rangle = \langle \nabla_{x_i}U \rangle \langle \nabla_{y_i}U\rangle=0$ and
$\langle \nabla_{x_i}U \nabla_{x_i}U\rangle=v_0^2$, we obtain that $\mathbf{\mathcal{D}}_{ij} \approx \tilde{\mathbf{\mathcal{D}}}_{ij}$, proving that, at very high density in the large persistence regime, the ABP velocity dynamics is well approximated by the AOUP dynamics.
Even with this simplification, 
an exact solution is not known, and we shall employ an approximation for the many-body velocity distribution~\cite{marconi2016velocity}:
\begin{equation}
\label{eq:approx_sol_wholepdf}
p(\{\mathbf{v}\} | \{\mathbf{x}\}) \propto \exp{\left(- \frac{1}{2v_0^2} \sum_{ij}\mathbf{v}_{i} \cdot\boldsymbol{\Gamma}_{ij} \cdot \mathbf{v}_j  \right)} \,.
\end{equation}
$p(\{\mathbf{v}\} | \{\mathbf{x}\})$ is a multivariate Gaussian coupling the whole set of velocities through the space dependent matrix $\boldsymbol{\Gamma}_{ij}$.
We remark that the prediction~\eqref{eq:approx_sol_wholepdf} is not the exact solution of the Fokker-Planck equation associated with the AOUP interacting dynamics, but it is a suitable approximation which works in the large persistent regime.
This prediction has been tested in several cases, even under the action of external potentials~\cite{caprini2018activeescape, caprini2019activityinduced}.

In the active solid phase, the matrix $\boldsymbol{\Gamma}_{ij}$, which depends on particles' relative positions, simplifies due to the hexagonal structure and to the short-range nature of the interaction potential.
Thus, the conditional probability distribution of $\mathbf{v}_i$ (i.e. knowing the velocity of the other particles) is given by Eq.~\eqref{eq:approx_sol_wholepdf}
where the sum is restricted to the first six neighbors of the target particle.
Integrating out all the velocity degrees of freedom except $\mathbf{v}_i$, we still obtain a Gaussian distribution with zero average confirming the shape reported in Fig.~\eqref{fig:velocitypdf} in the small persistence regime:
\begin{equation}
p(\mathbf{v}) \propto \exp{\left(- \frac{\beta}{2}\mathbf{v}^2  \right)}
\end{equation}
where $\beta$ is the variance of the distribution or the  inverse of the  kinetic temperature. 
Its exact expression as a function on the active force parameters will be derived in Sec.~\ref{Sec:kineticTemperature}.

\subsubsection{Small persistence-regime}

In the small persistence regime, the AOUP approximation is no longer valid, as numerically shown in Fig.~\eqref{fig:velocitypdf}.
Indeed, according to the AOUP model, the shape of $p(v_x, v_y)$ should always be Gaussian (with asymptotic corrections) at variance with our numerical results obtained with ABP simulations.
As shown in Fig.~\ref{fig:PotVel}, we  simplify the noise matrix assuming $v_0 \approx\langle|\mathbf{v}|\rangle \gg \langle|\mathbf{F}|\rangle$, obtaining
\begin{equation}
\label{eq:matrix_tausmall}
\begin{aligned}
\mathbf{\mathcal{D}}_{ij} &\approx \delta_{ij}\frac{1}{v_0^2} \begin{bmatrix} 
v_y^2 \quad& -v_y v_x  \quad&0\\
-v_y v_x  \quad& v_x^2 \quad&0 \\
0\quad&0\quad&0  
\end{bmatrix}\,.
\end{aligned}
\end{equation}
The Fokker-Planck equation~\eqref{eq:app_FokkerPlanck} with the matrix~\eqref{eq:matrix_tausmall}  (in the small $\tau$ limit) turns to be
\begin{equation}
\label{eq:app_FokkerPlanck_approx}
\begin{aligned}
\frac{\partial}{\partial t} p \approx& -\mathbf{v}_i \cdot \nabla_{\mathbf{x}_i} p + \frac{1}{\tau} \nabla_{\mathbf{v}_i} \cdot\left( \mathbf{v}_j p\right) \\
&+ \frac{\nabla_{\mathbf{x}_i} U}{\tau \gamma}\cdot\nabla_{\mathbf{v}_i} p + \frac{v_0^2}{\tau}\nabla_{\mathbf{v}_i} \nabla_{\mathbf{v}_j} \left[\mathcal{D}_{ij} p\right] \,.
\end{aligned}
\end{equation}
Neglecting the term $\propto \nabla_{\mathbf{x}_i} U $ because the forces almost cancel out in the solid phase (and their modulus is smaller than the velocity modulus, as shown in Fig.~\ref{fig:PotVel}), we can easily check that Eq.~\eqref{eq:app_FokkerPlanck_approx} admits a solution of the form:
\begin{equation}
\label{eq:analyticalresulst_smalltau}
p(v_x, v_y) \propto \exp{\left[- \frac{\alpha}{2}\left(|\mathbf{v}| - v_0 \right)^2\right]}  \,.
\end{equation}
The shape of Eq.~\eqref{eq:analyticalresulst_smalltau} corresponds to the Cartesian version of the velocity distribution shape numerically observed, i.e. Eq.~\eqref{eq:distribution_taularge}.

\section{The kinetic temperature}\label{Sec:kineticTemperature}

\begin{figure}[t]
\centering
\includegraphics[width=0.9\columnwidth,clip=true]{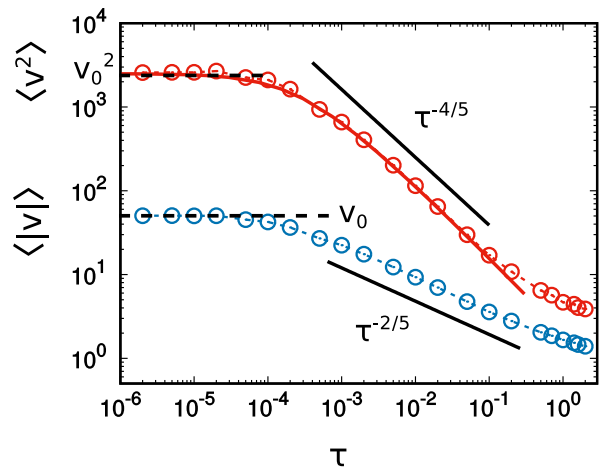}
\caption{
$\langle \mathbf{v}^2\rangle$ (red data) and $\langle |\mathbf{v}| \rangle$ (blue data) as a function of $\tau$. 
The colored dashed lines are plotted as a eye-guides, while the solid red line represents the theoretical prediction, Eq.~\eqref{eq:fourierspace_variance_v_cont}.
Numerical simulations are realized with $v_0=50$ and $\epsilon=\sigma=\gamma=1$.
}
\label{fig:vapred}
\end{figure}

In Fig.~\ref{fig:vapred}, we show the first two moments of the velocity modulus distribution as a function of $\tau$, namely $\langle |\mathbf{v}| \rangle$ and $\langle \mathbf{v}^2\rangle$.
The latter coincides by definition with the kinetic temperature of a system of ABP and, for this reason, we will denote $\langle \mathbf{v}^2\rangle$ simply as ``kinetic temperature''  in the rest of the paper.
For $\tau \lesssim 10^{-4}$, the system is in the small persistence regime [case i)] and both $\langle |\mathbf{v}| \rangle$ and $\langle \mathbf{v}^2\rangle$ are roughly constant with $\tau$, being approximatively $\langle |\mathbf{v}| \rangle\approx v_0$ and $\langle \mathbf{v}^2 \rangle  \approx v_0^2$.
We recall that, in this regime, the interparticle forces almost balance and the velocity displays the same statistical properties of the self-propulsion in such a way that the kinetic temperature does not display any $\tau$-dependence.
Upon increasing $\tau$, the values of $\langle |\mathbf{v}| \rangle$ and $\langle \mathbf{v}^2 \rangle$ monotonically decrease reaching very small values.
The more persistent is the particle motion, the slower it becomes and, as a consequence, the kinetic temperature decreases monotonically with $\tau$.
After a crossover regime occurring for $10^{-4}\lesssim\tau\lesssim10^{-3}$, a clear power-law scaling with $\tau$ appears for $10^{-3}\lesssim\tau\lesssim10^{-1}$ in both the moments. In particular, we have $\langle |\mathbf{v}| \rangle \sim v_0 (\gamma\tau)^{-2/5}$ and $\langle \mathbf{v}^2 \rangle\sim v_0^2 (\gamma\tau)^{-4/5}$, as clearly shown in Fig.~\ref{fig:vapred}.
The validity of these scalings ceases approximatively at $\tau=10^{-1}$, i.e. near the solid-hexatic transition.
Starting from this value of $\tau$, both $\langle |\mathbf{v}| \rangle$ and $\langle \mathbf{v}^2 \rangle$ decrease slower than $\sim\tau^{-2/5}$ and $\sim \tau^{-4/5}$ as the persistence time is increased without showing any clear power-law scaling with $\tau$.

In what follows, we develop an exact, analytical prediction valid in active the solid-state for the kinetic temperature which will explain the scaling with $\tau$ numerically observed shedding light also on the role of the other parameters.
Indeed, the periodicity of the almost-solid structure and, in particular, its hexagonal order suggests switching in the Fourier space to perform calculations~\cite{caprini2020hidden, caprini2020time}.
As reported in Appendix~\ref{app:appendixderivation}, the velocity correlation in the Fourier space reads:
\begin{equation}
\label{eq:vv_fourierspace}
\langle \hat{\mathbf{v}}_{\mathbf{q}} \cdot \hat{\mathbf{v}}_{-\mathbf{q}} \rangle  = 
  \frac{v_0^2}{1+\frac{\tau}{\gamma}\omega_\bq^2}\\
\end{equation}
where $\hat{\mathbf{v}}_{q}$ is the Fourier transform of the velocity vector $\mathbf{v}$ and $\mathbf{q}=(q_x, q_y)$ is a vector of the reciprocal Bravais lattice. The factor $\omega^2_{\mathbf{q}}$ has the following form:
\begin{flalign}
\label{eq:omega2_main}
\omega^2_{\mathbf{q}}=-2 K \Bigl[\cos(q_x \bar{r}) +2\cos\Bigl(\frac{1} {2} q_x \bar{r}\Bigr)\cos \Bigl(\frac{\sqrt 3} {2} q_y \bar{r}\Bigr)-3\Bigr] 
\end{flalign}
where the dimensional constant $K$ reads
\begin{equation}
2K =U''(\bar{r}) - \frac{U'(\bar{r})}{\bar{r}} \,.
\end{equation}
and $\bar{r}$ is the average distance between neighoring particles.
To obtain the variance of the velocity, we need to to go back to the real space calculating the inverse Fourier trasform in the origin: 
\begin{equation}
\langle \vv^2\rangle
=\frac{v_0^2}{N^2}  \sum_{\bq} \frac{1}{ (1+\frac{\tau}{\gamma}\omega_\bq^2)}  \,.
\end{equation}
Taking the continuum limit in the $\mathbf{q}$-sum and accounting for the periodicity of the lattice, we restrict the integral to the first Brillouin zone:
\begin{equation}
\label{eq:fourierspace_variance_v_cont}
\langle \vv^2\rangle
\approx v_0^2  \mathcal{I}\left[\frac{\tau}{\gamma}\right]  \,,
\end{equation}
where
\begin{equation}
 \mathcal{I}\left[\frac{\tau}{\gamma}\right] =  \frac{\bar{r}^2}{|\mathcal{B}|} \int_{\mathcal{B}} d{\mathbf{q}} \frac{1}{ (1+\frac{\tau}{\gamma}\omega_\bq^2)} \,,
\end{equation}
and $|\mathcal{B}|$ is the area of the Brillouin region associated with the hexagonal lattice.
We remark that it is not possible to approximate $\omega^2_{\mathbf{q}}$ for small $\mathbf{q}$ truncating at the quadratic order since the integral diverges at $\mathbf{q}=0$.
To the best of our knowledge, Eq.~\eqref{eq:fourierspace_variance_v_cont} is the first analytical expression for the kinetic temperature of interacting self-propelled particles that does not require  fitting parameters.
We observe that our expression increases quadratically with $v_0$ in agreement with previous results~\cite{petrelli2020effective} while the dependence on packing fraction and persistence time is contained in the integral $\mathcal{I}\left[\frac{\tau}{\gamma}\right]$.
As shown in Fig.~\ref{fig:vapred} (see the comparison between red points and the solid red line), Eq.~\eqref{eq:fourierspace_variance_v_cont} is in fair agreement with numerical data when the system attains solid configurations for $\tau\lesssim 10^{-1}$.
On one hand, $\mathcal{I}[0]\approx 1$ for small values of $\tau$, while, on the other hand, the numerical integration of $\mathcal{I}$ confirms both the crossover regime and the scaling $\sim (\tau/\gamma)^{-4/5}$ in the large persistence regime.
For $\tau\gtrsim 10^{-1}$, Eq.~\eqref{eq:fourierspace_variance_v_cont} underestimates the values of $\langle \mathbf{v}^2 \rangle$ with respect to numerical data
because, for these values of $\tau$, the structure of the system is no longer a solid without defects. Thus, a fundamental hypothesis behind the derivation of the prediction is violated and, thus, Eq.~\eqref{eq:fourierspace_variance_v_cont} is no longer valid.
In particular, it has been already shown that, in the proximity of defects, active particles have kinetic energies much larger than the ones in the absence of defects, as occurs in active solid configurations~\cite{caprini2020hidden}.
This is a clue to understanding why the decrease of $\langle \mathbf{v}^2\rangle$ with $\tau$ in active hexatic phases is slower than the decrease for active solids.
We remark that the integral $\mathcal{I}$ contains also the dependence on the packing fraction through the constant $K$ in Eq.\eqref{eq:omega2_main}.
Indeed, $K$ is mainly determined by the second derivative of the potential calculated at $\bar{r}$ that is uniquely fixed by the packing fraction in any homogeneous configurations.
Thus, the growth of $\phi$ induces the increase of the kinetic temperature through the non-linear derivatives of the function $U(\bar{x})$.
The explicit dependence on the potential shape is in agreement with previous studies based on temperature definitions derived in simpler cases, i.e. a one-dimensional particle confined through an external potential~\cite{marconi2017heat, caprini2019activityinduced}.


\subsection{Higher order moments and non-Gaussianity}

\begin{figure}[t]
\centering
\includegraphics[width=0.85\columnwidth,clip=true]{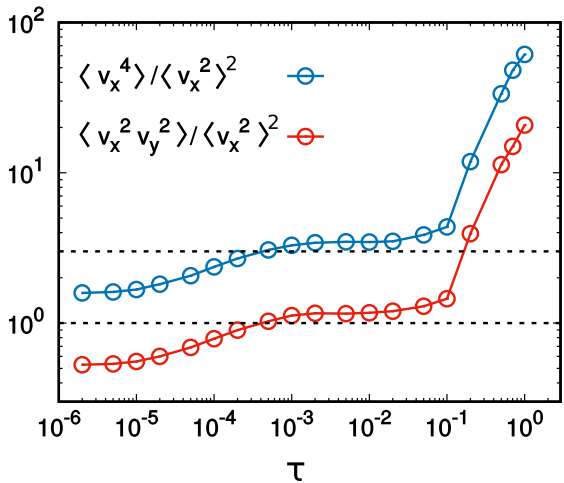}
\caption{
$\langle v^4_x\rangle/\langle v^2_x\rangle^2$ (blue data) and $\langle v^2_x v^2_y\rangle/\langle v^2_x\rangle^2$ (red data) as a function of $\tau$.
The colored solid lines are eye-guides while the dashed black lines are marked in correspondence of $\langle v^4_x\rangle/\langle v^2_x\rangle^2=3$ and $\langle v^2_x v^2_y\rangle/\langle v^2_x\rangle^2=1$, i.e. at the expected values for a velocity following the Gaussian statistics.
Numerical simulations are realized with $v_0=50$ and $\epsilon=\sigma=\gamma=1$.
}
\label{fig:v4}
\end{figure}

In spite of the fact that the equilibrium-like Gaussian prediction is a good approximation of the velocity distribution, at least in the large persistence regime, even from Fig.~\ref{fig:velocitypdf}~(d) clear deviations from the Gaussian theory 
are evident in its tails.
To get a quantitative analysis of the non-Gaussianity, we report the behavior of the higher-order moments in Fig.\ref{fig:v4}.
In particular, we show the kurtosis of the $v_x$-distribution, namely $\langle v^4_x \rangle/\langle v^2_x \rangle^2$.
This observable is rather small (around $\sim 1$) in the small-$\tau$ regime as a result of the non-Gaussianity of the distribution (Fig.~\ref{fig:velocitypdf}~(c)).
On the contrary, in the Gaussian-like regime, for $\tau \gtrsim 10^{-3}$, the kurtosis shows just small departures from the Gaussian prediction corresponding to $\langle v^4_x \rangle/\langle v^2_x \rangle^2 = 3$.
In particular, our numerical observations reveal that $\langle v^4_x \rangle/\langle v^2_x \rangle^2 \geq 3$, meaning that the tails of the distribution are a little fatter than the Gaussian prediction.
Finally, for larger values of $\tau$, i.e. for $\tau > 10^{-1}$ at the solid-hexatic transition point, the kurtosis abruptly increases and the system departs from the Gaussian-like regime.
As already mentioned, this is consistent with the occurrence of intermittency phenomena in the hexatic phase~\cite{caprini2020hidden} that manifest themselves also in high and non-Gaussian peaks in the time-trajectory of the single-particle kinetic energy.
As a further confirmation, a similar scenario, consistent with the observation regarding the kurtosis, occurs for the observable, $\langle v_x^2 v_y^2 \rangle /\langle v_x^2 \rangle^2$.
In particular, $\langle v_x^2 v_y^2 \rangle /\langle v_x^2 \rangle^2 \lesssim 1$ in the small persistence regime, $1$ for regime with $10^{-3}\leq\tau \leq 10^{-1}$ (as expected in Gaussian regimes) and $\gtrsim 1$ when the hexatic phase occurs.


\section{Discussion and Conclusions}

In this paper, we have studied the properties of the velocity of highly packed systems of self-propelled particles (active hexatic and solid phases) to understand the influence of the activity.
A transition from Mexican-hat-like velocity distribution (i.e.  peaked in the proximity of a circular crown with a radius larger than zero) to Gaussian-like velocity distribution 
is observed going from the small persistence to the large persistence time regime.
Analyzing the velocity dynamics, we derive suitable approximations to predict the functional form of the probability distribution function of the velocities in both these regimes.
Concerning the active solid, we derive by a Fourier-space method a theoretical expression
for the variance of the velocity distribution, giving the kinetic temperature of ABP in this phase.
Thus, on one hand, we have derived new approximate analytical results concerning the velocity distribution of ABP particles holding both near and far from equilibrium and, on the other hand, we have provided the analytical expression for the active kinetic temperature in the solid phase.

At least in homogeneous solid configurations, our analytical expression for the kinetic temperature increases quadratically with the speed of the self-propelled particles (that is proportional to the Peclet number).
This quadratic scaling has been also observed by means of different definitions of temperature, such as the active effective temperature for homogeneous configurations~\cite{loi2011non, suma2014dynamics}.
In those cases, our kinetic temperature displays a monotonic decrease as a function of the packing fraction as also for the effective temperature~\cite{petrelli2020effective}.
Our results show that the kinetic temperature contains a strong dependence on the shape of the interacting potential in agreement with another temperature definition obtained in the case of single-particle confined through external potentials~\cite{marconi2017heat, caprini2019activityinduced}.
Thus, the concavity of the potential plays a fundamental role not only for confined non-interacting active particles~\cite{caprini2019activeescape} but also for interacting systems, being relevant to determine the velocity variance (and, thus, the kinetic temperature) in both cases.
While the kinetic temperature does not show a dependence on $\tau$ in the small persistence regime, a power-law decay with $\tau$ is numerically observed and theoretically predicted in the large persistence regime.
We remark that our predictions are valid in the active solid-state while does not work where orientational and/or positional orders are broken (i.e. active hexatic and liquid state, respectively). In those cases, the occurrence of large non-Gaussianity and intermittency phenomena in the time-trajectory of the kinetic energy are consistent with the failure of the theoretical predictions~\cite{caprini2020hidden}.

\subsection*{Acknowledgements}
LC thanks I. Petrelli, M. Cencini and A. Puglisi for fruitful discussions.
LC and UMBM acknowledge support from the MIUR PRIN 2017 project 201798CZLJ.

\appendix 
\section{The kinetic temperature of active particles in the solid-state}\label{app:appendixderivation}

To develop a prediction for the kinetic temperature of self-propelled particles in the active solid state, we shall employ two approximations to simplify the dynamics, Eq.~\eqref{eq:x_dynamics_ABP}. i) The dynamics of each component of $\mathbf{f}^a_i$ is replaced by independent Ornstein-Uhlenbeck processes with equivalent persistence time,  $\tau=1/D_r$, and variance $v_0^2$ in such a way that $\langle|\mathbf{f}^a|\rangle=v_0$ consistently with the ABP model. 
ii) Each particle oscillates around a node of a hexagonal lattice so that the 
total interparticle potential is approximated as the sum of quadratic terms.
With these two assumptions, the original dynamics, Eq.~\eqref{eq:x_dynamics_ABP}, becomes:
\begin{flalign}
&\dot \bR_i(t) =  \mathbf{f}_i^a(t)  
-  \sum_j^{n.n}\frac{\nabla_i U(|\bR_j-\bR_i|)}{\gamma} 
 \label{dynamicequation0}\\
&\tau\dot{\mathbf{f}}_i^a(t) =-\mathbf{f}_i^a(t)+ v_0\sqrt{2 \tau}\, \xxi_i(t) \,,
\end{flalign}
where the sum involves the nearest neighbors
of the lattice node $i$, and the symbol $\nabla_i$ is the gradient with respect to $\bR_i$.
Introducing the displacement $\bu_i$ of the particle $i$ with respect to its equilibrium position, $\bRz_i$, namely
\begin{equation}
\bu_i=\bR_i-\bRz_i \,,
\end{equation}
we obtain
\begin{flalign}
 \label{dynamicequation2}
&\dot \bu_i(t) =  \mathbf{f}_i^a(t)  
+\frac{K}{\gamma} \sum_j^{n.n} (\bu_j-\bu_i) \\
&\tau\dot{\mathbf{f}}_i^a(t) =-\mathbf{f}_i^a(t)+ v_0\sqrt{2 \tau}\, \xxi_i(t) \,,
\end{flalign}
%
being $K$ the strength of the potential in the harmonic approximation, i.e. $U\approx\frac{K}{2} (\bu_j-\bu_i)^2$, that explicitly reads:
$$
2 K = \left(U''(\bar{r}) + \frac{U'(\bar{r})}{\bar{r}} \right) \,,
$$
where $\bar{r}$ is the lattice constant.
Because of the linearity of the system, it is useful to switch 
to normal coordinates, in the Fourier space representation:
\begin{eqnarray}
&&
 \hat \bu_{\bq}=\frac{1}{ N}\sum_i   \bu_i\,  e^{-i \bq\cdot  \bRz_i }
\\&&
\hat \eeta_{\bq}=\frac{1}{ N}\sum_i   \eeta_i\,  e^{-i \bq\cdot  \bRz_i } \,,
\label{fourierrepresentation}
\end{eqnarray}
where $\hat \bu_{\bq}$ and $\hat \eeta_{\bq}$ are the Fourier transform of $\bu$ and $\mathbf{f}^a$, respectively.
The dynamics in the Fourier Space reads:
\begin{flalign}
\label{dynamicequation4}
&\frac{d}{dt}\hat \bu_\bq(t) =-\frac{\omega^2_\bq}{\gamma} \hat \bu_\bq(t) + \hat \eeta_\bq \\
&\tau\frac{d}{dt}\hat{\eeta}_\bq(t) =- \hat \eeta_\bq+v_0\sqrt{ 2 \tau} \,\hat \xxi_\bq \,,
\end{flalign}
where
\begin{flalign}
\label{eq:app_omegaq}
\omega_\bq^2&=-2 K \Bigl[\cos(q_x \bar{r}) +2\cos\Bigl(\frac{1} {2} q_x \bar{r}\Bigr)\cos \Bigl(\frac{\sqrt 3} {2} q_y \bar{r}\Bigr)-3\Bigr] \nonumber
\end{flalign}
where $\mathbf{q}=(q_x, q_y)$ are vectors of the reciprocal Bravais lattice.
Defining  $\hat{\vv}_\bq$ as the Fourier transform of the velocity $\mathbf{v}$, that satisfies $\hat{\vv}_\bq = \frac{d}{dt}\hat \bu_\bq$, 
we can easily calculate the steady-state equal time correlations in the Fourier space, that is:
 \begin{equation}
\label{eq:app_vvfourier}
\langle \hat{\vv}_\bq \cdot \hat{\vv}_{-\bq}  \rangle=
  \frac{2 v_0^2}{1+\frac{\tau}{\gamma}\omega_\bq^2} \,.
\end{equation}
Eq.~\eqref{eq:app_vvfourier} is the final expression for the spatial velocity correlation in the Fourier space and corresponds to Eq.~\eqref{eq:vv_fourierspace}.

\bibliographystyle{apsrev4-1}

\bibliography{bib.bib}

\end{document}